\documentclass[twocolumn,preprintnumbers,amsmath,amssymb]{revtex4}

\usepackage{graphicx}
\usepackage{dcolumn}
\usepackage{bm}
\usepackage{verbatim}

\newcommand{\bra}[1]{\ensuremath{\langle #1 \vert}}
\newcommand{\ket}[1]{\ensuremath{\vert #1  \rangle}}
\newcommand{\braket}[2]{\ensuremath{\langle  #1 \vert #2  \rangle}}

\renewcommand{\b}[1]{\ensuremath{\mathbf{#1}}}

\newcommand{\sgn}{\ensuremath{\text{sgn}}}
\newcommand{\VMC}{\ensuremath{\text{VMC}}}
\newcommand{\DMC}{\ensuremath{\text{DMC}}}

\newcommand{\HF}{\ensuremath{\text{HF}}}
\newcommand{\ZV}{\ensuremath{\text{ZV}}}
\newcommand{\ZB}{\ensuremath{\text{ZB}}}
\newcommand{\ZVZB}{\ensuremath{\text{ZVZB}}}
\newcommand{\FN}{\ensuremath{\text{FN}}}
\newcommand{\hybrid}{\ensuremath{\text{hybrid}}}

\newcommand{\opt}{\ensuremath{\text{opt}}}

\renewcommand{\l}{\ensuremath{\lambda}}

\newcommand{\z}{\ensuremath{\zeta}}
\newcommand{\la}{\ensuremath{\left\langle}}
\newcommand{\ra}{\ensuremath{\right\rangle}}
\newcommand{\C}{\ensuremath{\text{C}}}
\newcommand{\N}{\ensuremath{\text{N}}}

\newcommand{\OS}{\ensuremath{\text{OS}}}
\renewcommand{\SS}{\ensuremath{\text{SS}}}

\begin{document}

\title{Zero-variance zero-bias quantum Monte Carlo estimators of the spherically and system averaged pair density}

\author{Julien Toulouse}
\email{toulouse@tc.cornell.edu}
\affiliation{
Cornell Theory Center, Cornell University, Ithaca, New York 14853, USA.
}
\author{Roland Assaraf}
\email{assaraf@lct.jussieu.fr}
\affiliation{
Laboratoire de Chimie Th\'eorique, Universit\'e Pierre et Marie Curie and Centre National de la Recherche Scientifique, 75005 Paris, France.
}
\author{C. J. Umrigar}
\email{cyrus@tc.cornell.edu}
\affiliation{
Cornell Theory Center and Laboratory of Atomic and Solid State Physics, Cornell University, Ithaca, New York 14853, USA.
}

\date{\today}

\begin{abstract}
We construct improved quantum Monte Carlo estimators for the spherically- and system-averaged electron pair density
(i.e. the probability density of finding two electrons separated by a relative distance $u$),
also known as the spherically-averaged electron position intracule density $I(u)$,
using the general zero-variance zero-bias principle for observables,
introduced by Assaraf and Caffarel.
The calculation of $I(u)$ is made vastly more efficient by replacing the average of the local delta-function
operator by the average of a smooth non-local operator that has several orders of magnitude smaller variance.
These new estimators also reduce the systematic error (or bias) of the intracule density due to the
approximate trial wave function.
Used in combination with the optimization of an increasing number of parameters in
trial Jastrow-Slater wave functions, they allow one to obtain well converged correlated intracule densities for atoms and molecules.
These ideas can be applied to calculating any pair-correlation function in classical or quantum Monte Carlo calculations.
\end{abstract}

\maketitle

\section{Introduction}
\label{sec:intro}

Two-electron distribution functions occupy an important place in electronic structure theory between the simplicity
of one-electron densities and the complexity of the many-electron wave function.
In particular, the system-averaged electron pair density,
the probability density of finding two electrons separated by the relative position vector $\b{u}$,
also known as the electron position intracule density $I(\b{u})$,
plays an important role in qualitative and quantitative descriptions of electronic systems.
Position intracule densities have been extensively used to analyze shell structure, electron correlation, Hund's rules and chemical bonding (see, e.g., Refs.~\onlinecite{CouNei-PPSL-61,LesKra-JCP-66,Koh-JCP-72,Kat-PRA-72,BoyCou-JPB-73,CooPou-TCA-77,ThaTriSmi-IJQC-84,RegTha-JPB-84,ShaTha-JPB-84,UgaBoy-IJQC-85,BoySarUga-JPB-88,SarUgaBoy-JPB-90,SarDomAguUga-JCP-92,WanTriSmi-JCP-92,CanBoyTha-JCP-93,WanSmi-IJQC-94,AriPorBueGal-JPB-95,MeyMulSch-JMS-96,CioLiu-JCP-96b,FraDurMes-JCP-97,MatKogRomDeh-PRA-98,SarGalBue-JCP-98,GalBueSar-JCP-99,CioLiu-JCP-99,GalBueSar-PRA-00,GalBueSar-JCP-02,GilOneBes-TCA-03,MerValUga-INC-03,GalBueSar-JCP-03,GalBueSar-JCP-05,GalBueSar-JCP-05a}). Density functional theory-like approaches have been proposed based on the position intracule density~\cite{Kog-JCP-90,GorSav-PRA-05,GorSav-PM-06,GorSav-JJJ-XX,Nag-JCP-06} or on the closely-related Wigner intracule density~\cite{GilCriOneBes-PCCP-06,GilOne-JCP-05,Bes-JCP-06}.

Position intracule densities have been extracted from experimental X-ray scattering intensities for small atoms and molecules~\cite{ThaTriSmi-PRA-84,WanTriSmi-JCP-94,WatKamYamUdaMul-MP-04}. They have been calculated in the Hartree-Fock (HF) approximation for systems ranging from small atoms to large molecules~\cite{ThaTriSmi-IJQC-84,SarUgaBoy-JPB-90,WanSmi-IJQC-94,CioLiu-JCP-96a,CioLiu-JCP-96b,MatKogRomDeh-PRA-98,KogMatDehTha-JCP-99,LeeGil-CPL-99,GilLeeNarAda-JMS-00,GilOneBes-TCA-03}. Calculations using common quantum chemistry correlated methods such as second-order M{\o}ller-Plesset perturbation theory, multi-configurational self-consistent-field (MCSCF) and configuration interaction approaches have been limited to atoms and small molecules~\cite{CioLiu-JCP-98,CioLiu-JCP-99,Kog-JCP-02,CooPou-TCA-77,BoySarUga-JPB-88,WanTriSmi-JCP-92,MeyMulSch-JMS-96,MerValUga-INC-03}. Very accurate calculations using Hylleraas-type explicitly-correlated wave functions have been done only for the helium and lithium isoelectronic series~\cite{CouNei-PPSL-61,BoyCou-JPB-73,ThaSmi-JCP-77,RegTha-JPB-84,Tha-INC-87,CanBoyTha-JCP-93,DreKin-JCP-94,AriPorBueGal-JPB-95,CioSteTanUmr-JCP-95,Tha-CPL-03}. Variational Monte Carlo (VMC) calculations using Jastrow-Slater wave functions have been used to compute correlated position intracule densities for atoms from helium to neon and some of their isoelectronic series~\cite{CioSteTanUmr-JCP-95,SarGalBue-JCP-98,GalBueSar-JCP-99,GalBueSar-PRA-00,GalBueSar-JCP-02,GalBueSar-CPL-03,GalBueSar-CPL-03a,GalBueSar-JCP-03,GalBueSar-JCP-05,GalBueSar-JCP-05a}.
In this paper, we show that the calculation of position intracule densities using quantum Monte Carlo (QMC) methods can
be made much more accurate and efficient, opening new possibilities of investigation.

The position intracule density associated with an $N$-electron (real) wave function $\Psi(\b{R})$, where
$\b{R}=(\b{r}_1,\b{r}_2,...,\b{r}_N)$ is the $3N-$dimensional vector of electron coordinates (ignoring spin for now),
is defined as the quantum-mechanical average of the delta-function operator $\delta(\b{r}_{ij} - \b{u})$
\begin{eqnarray}
I(\b{u}) = \frac{1}{2} \sum_{i \not= j} \int d\b{R} \, \Psi(\b{R})^2 \, \delta(\b{r}_{ij} - \b{u}),
\end{eqnarray}
where $\b{r}_{ij}=\b{r}_{j}-\b{r}_{i}$ and the sum is over all electron pairs, and its spherical average is
\begin{eqnarray}
I(u) = \int \frac{d\Omega_{\b{u}}}{4\pi} \, I(\b{u}).
\end{eqnarray}
Among the most important properties of $I(u)$ are the normalization sum rule (giving the total number of electron pairs)
\begin{eqnarray}
\int_0^\infty du \, 4\pi u^2 \, I(u) = \frac{N(N-1)}{2},
\end{eqnarray}
the electron-electron cusp condition~\cite{ThaSmi-CPL-76}
\begin{eqnarray}
\frac{dI(u)}{du} \Biggl|_{u=0}=I(0),
\label{cusp_cond}
\end{eqnarray}
and, for finite systems, the exponential decay at large $u$ determined by the spherically averaged one-electron density $n(u)$ evaluated at the distance $u$ from the chosen origin
\begin{eqnarray}
I(u) \underset{u \to \infty}{\sim} \frac{(N-1)}{2} n(u)  \underset{u \to \infty}{\propto} e^{-2 \sqrt{2 I} \, u},
\label{Idecay}
\end{eqnarray}
where $I$ is the vertical ionization energy (see Refs.~\onlinecite{LevPerSah-PRA-84,AlmBar-PRB-85,ErnBurPer-JCP-96}). The moments of the radial intracule density $M_k =\int_0^\infty du \, 4\pi u^{2+k} \, I(u)$ are related to physical observables~\cite{KogMat-JCP-01}, in particular $M_{-1}$ is just the electron-electron Coulomb interaction energy
\begin{eqnarray}
W_{ee} = M_{-1} =  \int_0^\infty du \, 4\pi u \, I(u).
\label{Wee}
\end{eqnarray}

In standard correlated methods based on an expansion of the wave function in Slater determinants, the important short-range part of the position intracule density converges very slowly with the one-electron and many-electron basis. The advantage of employing QMC methods~\cite{FouMitNeeRaj-RMP-01} lies in the possibility of using compact, explicitly-correlated wave functions which are able to describe properly the short-range part of $I(u)$. However, the problem in this approach is that one calculates the average of a delta-function operator which has an \textit{infinite} variance.
As for other probability densities, the standard procedure in QMC approaches simply consists of counting the number of electron pairs separated by the distance $u$ within $\Delta u$ encountered in the Monte Carlo run. More precisely, e.g. in VMC calculations, $I(u)$ is estimated as the statistical average
\begin{eqnarray}
I^{\text{histo}}(u) = \la I_{L}^{\text{histo}}(u, \b{R}) \ra_{\Psi^2},
\end{eqnarray}
over the trial wave function density $\Psi(\b{R})^2$, of the local ``histogram'' estimator $I_{L}^{\text{histo}}(u, \b{R})$
\begin{eqnarray}
I_{L}^{\text{histo}}(u,\b{R}) = \frac{1}{2} \sum_{i \not= j} \frac{1_{[u - \Delta u /2 \, , \, u + \Delta u /2[}(r_{ij})}{4 \pi u^2 \Delta u},
\label{esthisto}
\end{eqnarray}
where $1_{[a,b[}(r)=1$ if $r \in [a,b[$ and $0$ otherwise. Here and in the following, $\la f(\b{R}) \ra_{\Psi^2} = (1/M) \sum_{k=1}^{M} f(\b{R}_k)$ designates the statistical average of the local quantity $f(\b{R})$ over $M$ configurations $\b{R}_k$ sampled from $\Psi(\b{R})^2$. For $u \not = 0$, this histogram estimator has a finite but \textit{large variance} for small $u$ or small $\Delta u$. Although the use of importance sampling can help decrease the variance~\cite{LanRotVrb-JCP-97,SarGalBue-JCP-98,SarGalBue-CPC-99}, the calculation of position intracule densities in QMC remains very inefficient: very long runs have to be performed to reach an acceptably small statistical uncertainty. Moreover, the histogram estimator has a \textit{discretization error}: even in the limit of an infinite sample $M \to \infty$, $I^{\text{histo}}(u)$ remains only an approximation of first order in $\Delta u$ to the position intracule density $I(u)$ of the wave function $\Psi(\b{R})$.
Note however that it is possible to greatly reduce the discretization error by choosing a flexible analytic form for $I(u)$ that obeys known conditions (such as the cusp condition of Eq.~(\ref{cusp_cond})) and fitting in each interval $\Delta u$ the integral of $I(u)$ (instead of $I(u)$) to the computed data. This approach has been used to calculate radial electron densities for atoms~\cite{FilGonUmr-INC-96} and spherically-averaged intracule densities~\cite{CioSteTanUmr-JCP-95} but the method is not as effective for multidimensional densities.

In this work, we develop improved QMC estimators for the spherically averaged position intracule density based on the \textit{zero-variance zero-bias} (ZVZB) principle for observables introduced by Assaraf and Caffarel for calculations of electronic forces~\cite{AssCaf-PRL-99,AssCaf-JCP-00,AssCaf-JCP-03}, which has been recently applied to calculations of one-electron densities~\cite{AssCafSce-PRE-07}. These new ZVZB estimators of $I(u)$ have variances several orders of magnitude smaller than the ones obtained with the standard estimator of Eq.~(\ref{esthisto}) and thus dramatically increase the efficiency of these calculations. Moreover, these estimators do not suffer from any discretization error and can even reduce the systematic error due to the approximate trial wave function. Like related techniques proposed for calculations of one-electron or two-electron densities using deterministic \textit{ab initio} methods~\cite{HilSucFei-PRA-78,SucDra-PRA-79,Kat-PRA-80,Tri-JPB-80,Har-IJQC-80,Dra-JPB-81,Har-IJQC-85,Har-IJQC-86,MomShi-JCP-87,Ish-CPL-89,ChaCio-JCP-94,CioSteTanUmr-JCP-95,LiuParNag-PRA-95,RasChi-JCP-96a,RasChi-JCP-96b,WanSchSmi-PRA-00}, these estimators replace the average of the local delta-function operator by the average of an operator which is nonlocal in real space. They can be viewed as a generalization of the improved QMC estimators previously proposed for computations of averages of probability densities at particle coalescences~\cite{VrbDepRot-JCP-88,LanRotVrb-JCP-97,BreMelMor-PRA-98,AleCol-JMS-99}.

Moreover, we make use in this work of the recently-developed linear energy minimization method~\cite{TouUmr-JCP-07,UmrTouFilSorHen-PRL-07} to finely optimize the Jastrow parameters, the configuration state functions (CSFs) coefficients and the orbital coefficients of our trial Jastrow-Slater wave functions. Optimization of the determinantal part of the wave function with an increasing number of CSFs allows us to obtain a systematic improvement of wave functions.
This provides a practical route for calculating intracule densities in VMC and fixed-node (FN) diffusion Monte Carlo (DMC) with progressively
smaller systematic errors.

The paper is organized as follows. In Sec.~\ref{sec:review_zvzb}, we review the principle of zero-variance zero-bias improved estimators for an arbitrary observable in QMC. In Sec.~\ref{sec:zvzb_intracule}, we give improved estimators for the case of the position intracule density. Sec.~\ref{sec:compdetails} contains computational details of the calculations, and Sec.~\ref{sec:results} discusses results for the He and C atoms and for the C$_2$ and N$_2$ molecules to illustrate the technique. Sec.~\ref{sec:conclusion} contains our conclusions.

Hartree atomic units are used throughout this work.

\section{Zero-variance zero-bias improved estimators}
\label{sec:review_zvzb}

We review here the principle of the zero-variance zero-bias improved QMC estimators for an arbitrary observable that does not commute with the Hamiltonian, developed in Refs.~\onlinecite{AssCaf-PRL-99,AssCaf-JCP-00,AssCaf-JCP-03}. Throughout this section, all the averages $\langle \cdots \rangle_{\Psi^2}$ are considered in the limit of an infinite Monte Carlo (MC) sample $M \to \infty$.

\subsection{Estimators in variational Monte Carlo}

In VMC, the exact energy $E_0=\bra{\Psi_0} \hat{H} \ket{\Psi_0}/\braket{\Psi_0}{\Psi_0}$ of some exact eigenfunction $\Psi_0$ of the electronic Hamiltonian $\hat{H}$ is estimated by the statistical average of the local energy $E_L(\b{R})= \bra{\b{R}} \hat{H} \ket{\Psi}/\Psi(\b{R})$
\begin{eqnarray}
E = \la E_L(\b{R}) \ra_{\Psi^2},
\end{eqnarray}
using an approximate trial wave function $\Psi(\b{R})$. The systematic error (or bias) of this estimator $\delta E = E - E_0$ and its variance $\sigma^2 \left(E_L\right) = \langle \left(E_L(\b{R}) -E\right)^2 \rangle_{\Psi^2}$, whose square root is proportional to the statistical uncertainty, both \textit{vanish quadratically} as a function of the error in the trial wave function $|\delta \Psi|=|\Psi - \Psi_0|$ (where $|\cdots|$ designates the Hilbert space norm)
\begin{subequations}
\begin{eqnarray}
\delta E = {\cal O}(|\delta \Psi|^2),
\end{eqnarray}
\begin{eqnarray}
\sigma^2  \left( E_L \right) = {\cal O}(|\delta \Psi|^2),
\end{eqnarray}
\label{Ezvzb}
\end{subequations}
which is referred to as the quadratic zero-variance zero-bias property of the local energy. This is easily shown by writing $\Psi = \Psi_0 + \delta \Psi$ in the expressions of the average and the variance, and expanding to second-order in $\delta \Psi$.

Similarly, the exact expectation value of an arbitrary observable $O_0 =\bra{\Psi_0} \hat{O} \ket{\Psi_0}/\braket{\Psi_0}{\Psi_0}$ can be estimated by the statistical average of the local observable $O_{L}(\b{R}) = \bra{\b{R}} \hat{O} \ket{\Psi}/\Psi(\b{R})$
\begin{eqnarray}
O = \la O_L(\b{R}) \ra_{\Psi^2},
\end{eqnarray}
but, as $\Psi_0$ is generally not an eigenstate of $\hat{O}$, the systematic error of this estimator $\delta O = O - O_0$ vanishes only linearly with $|\delta \Psi|$, while its variance $\sigma^2 \left(O_L\right) = \langle \left( O_L(\b{R}) - O \right)^2 \rangle_{\Psi^2}$ generally does not even vanish with $|\delta \Psi|$
\begin{subequations}
\begin{eqnarray}
\delta O = {\cal O}(|\delta \Psi|),
\label{deltaO}
\end{eqnarray}
\begin{eqnarray}
\sigma^2  \left( O_L \right) = {\cal O}(1),
\label{sigma2OL}
\end{eqnarray}
\end{subequations}
which often makes the calculations of observables inaccurate and inefficient.

However, Assaraf and Caffarel~\cite{AssCaf-JCP-03} have pointed out that the quadratic zero-variance zero-bias property of the energy can be extended to an arbitrary observable by expressing it as an energy derivative. This is based on the Hellmann-Feynman theorem which states that, if one defines the $\l$-dependent Hamiltonian
\begin{equation}
\hat{H}^\l = \hat{H} + \lambda \, \hat{O},
\end{equation}
with some associated exact $\l$-dependent eigenfunction
\begin{equation}
\Psi_0^\l = \Psi_0 + \lambda \, \Psi_0' + \cdots,
\end{equation}
where $\Psi_0'= \left( d \Psi_0^\l/d\l \right)_{\l=0}$ and corresponding exact $\l$-dependent energy $E_0^\l = \bra{\Psi_0^\l} \hat{H}^\l \ket{\Psi_0^\l} / \braket{\Psi_0^\l}{\Psi_0^\l}$, then the exact value of the observable is given by the derivative of the energy with respect to $\l$ at $\l=0$: $O_0 = \left( d E_0^\l/d\l \right)_{\l=0}$. Introducing an approximate $\l$-dependent trial wave function
\begin{equation}
\Psi^\l = \Psi + \lambda \, \Psi' + \cdots,
\end{equation}
where $\Psi'=\left( d \Psi^\l/d\l \right)_{\l=0}$, the $\l$-dependent energy can be estimated as the statistical average of the $\l$-dependent local energy $E_L^\l(\b{R})= \bra{\b{R}} \hat{H}^\l \ket{\Psi^\l}/\Psi^\l(\b{R})$
\begin{equation}
E^\l = \la E_{L}^{\l}(\b{R})  \ra_{{\Psi^{\l}}^2},
\end{equation}
over the probability density $\Psi^\l(\b{R})^2$, and the Hellmann-Feynman theorem suggests now to define a \textit{zero-variance zero-bias} (ZVZB) estimator for the observable as the derivative of $E^\l$ with respect to $\l$ at $\l=0$
 \begin{eqnarray}
O^{\ZVZB} &=& \la O_{L}^{\ZVZB}(\b{R}) \ra_{\Psi^2}
\nonumber\\
&=& \left( \frac{d E^\l}{d\l} \right)_{\l=0}
\nonumber\\
&=&  \la  O_{L}(\b{R}) \ra_{\Psi^2} + \la  \Delta O_{L}^{\ZV}(\b{R}) \ra_{\Psi^2}
\nonumber\\
&&+ \la  \Delta O_{L}^{\ZB}(\b{R}) \ra_{\Psi^2},
\end{eqnarray}
where the \textit{zero-variance} (ZV) contribution
\begin{eqnarray}
\Delta O_{L}^{\ZV}(\b{R}) &=& \left[ \frac{\bra{\b{R}} \hat{H} \ket{\Psi'}}{\Psi'(\b{R})} -E_L(\b{R}) \right] \frac{\Psi'(\b{R})}{\Psi(\b{R})},
\label{DOLZV}
\end{eqnarray}
does not contribute to the average, $\la \Delta O_{L}^{\ZV}(\b{R}) \ra_{\Psi^2}=0$ (in the limit of an infinite MC sample $M\to\infty$, from the Hermiticity of the Hamiltonian) but can decrease the variance, and the \textit{zero-bias} (ZB) contribution
\begin{eqnarray}
\Delta O_{L}^{\ZB}(\b{R}) = 2 \left[ E_L(\b{R}) - E \right] \frac{\Psi'(\b{R})}{\Psi(\b{R})},
\label{DOLZB}
\end{eqnarray}
usually has little effect on the variance but can decrease the systematic error. The average of this ZB term vanishes if the trial wave function $\Psi$ is an exact eigenstate or, more generally, if the energy $E$ is stationary with respect to an infinitesimal variation of the trial wave function along the derivative $\Psi'$: $\Psi \to \Psi + \epsilon \, \Psi'$. For the special case of the calculation of the derivatives of the energy with respect to the nuclear coordinates, the average of the ZB term is known in the electronic structure literature as the Pulay contribution~\cite{Pul-MP-69}.

If one wants simply to compute an observable for a given trial wave function without changing the average, the ZB correction has to be omitted. In this case, the ZV estimator $O_{L}^{\ZV}(\b{R}) = O_{L}(\b{R}) + \Delta O_{L}^{\ZV}(\b{R})$ has a systematic error that still vanishes linearly with $|\delta \Psi|$ but a variance that now vanishes \textit{quadratically} with $|\delta \Psi|$ and the error in its derivative $|\delta \Psi'| = |\Psi' - \Psi_0'|$
\begin{subequations}
\begin{eqnarray}
\delta O^{\ZV} = {\cal O}(|\delta \Psi|),
\label{deltaOZV}
\end{eqnarray}
\begin{eqnarray}
\sigma^2  \left( O_{L}^{\ZV} \right) = {\cal O}(|\delta \Psi|^2 + |\delta \Psi'|^2 + |\delta \Psi|\,|\delta \Psi'|).
\label{sigma2OZV}
\end{eqnarray}
\end{subequations}
If one also includes the ZB correction, then both the systematic error and the variance of the ZVZB estimator $O_{L}^{\ZVZB}(\b{R})$ vanish \textit{quadratically} with $|\delta \Psi|$ and $|\delta \Psi'|$
\begin{subequations}
\begin{eqnarray}
\delta O^{\ZVZB} = {\cal O}(|\delta \Psi|^2 + |\delta \Psi|\,|\delta \Psi'|),
\end{eqnarray}
\begin{eqnarray}
\sigma^2  \left( O_{L}^{\ZVZB} \right) = {\cal O}(|\delta \Psi|^2 + |\delta \Psi'|^2 + |\delta \Psi|\,|\delta \Psi'|).
\label{sigma2OZVZB}
\end{eqnarray}
\end{subequations}
This is easily shown by writing $\Psi = \Psi_0 + \delta \Psi$ and $\Psi' = \Psi_0' + \delta \Psi'$ in the expressions of the average and the variance, and expanding with respect to $\delta \Psi$ and $\delta \Psi'$. One can verify that the exact cancellation of the dominant contributions to the systematic error and the variance of Eqs.~(\ref{deltaO}) and~(\ref{sigma2OL}) by the ZV and ZB corrections is a direct consequence of the relationship between the exact wave function $\Psi_0$ and its derivative $\Psi_0'$ in first-order perturbation theory: $(\hat{H} -E_0) \ket{\Psi_0'} = - (\hat{O} -O_0) \ket{\Psi_0}$.

Thus, the ZVZB estimators permit in principle accurate and efficient calculations for observables, provided that a good approximation for the derivative $\Psi_0'$ is available. We note that optimizing the parameters $\b{p}$ in the trial wave function $\Psi(\b{p})$ by energy minimization presents the advantage of decreasing the sensitivity of the systematic error to the quality of the approximate derivative $\Psi'$, since in this case $\Psi'$ only needs to be a good approximation of $\Psi_0'$ in the orthogonal complement of the space spanned by the wave function derivatives with respect to the parameters at the optimal parameter values $\b{p}_\opt$, i.e. $\Psi_i = \left( \partial \Psi(\b{p})/\partial p_i \right)_{\b{p}=\b{p}_\opt}$. Indeed, the contribution to the average of the ZB term of any component of $\Psi'$ along $\Psi_i$ is proportional to the energy gradient with respect to parameter $p_i$ and thus vanishes at the stationary point: $\left\langle 2 [ E_L(\b{R}) - E ]  \Psi_i (\b{R})/\Psi(\b{R}) \right\rangle_{\Psi^2} = \left( \partial E/\partial p_i \right)_{\b{p}=\b{p}_\opt}=0$. This point is at the origin of the improved accuracy in the calculations of forces in QMC in Refs.~\onlinecite{CasMelRap-JCP-03,LeeMelRap-JCP-05}.

\subsection{Estimators in diffusion Monte Carlo}

DMC calculations within the FN approximation go beyond VMC calculations by replacing the VMC distribution $\Psi(\b{R})^2$ by the more accurate mixed FN-DMC distribution $\Psi_\FN(\b{R}) \Psi(\b{R})$ in statistical averages, i.e. $\la \cdots \ra_{\Psi^2} \to \la \cdots \ra_{\Psi_\FN \Psi}$. All the DMC variances have the same lowest order behaviors with respect to the error in the trial wave function as in VMC.
The systematic error of the energy now vanishes as the product of the error in the trial wave function $|\delta \Psi|$ and the error in the FN wave function $|\delta \Psi_\FN|=|\Psi_\FN - \Psi_0|$
\begin{eqnarray}
\delta E = {\cal O}(|\delta \Psi|\,|\delta \Psi_\FN|).
\end{eqnarray}
For an arbitrary observable, the systematic error still vanishes only linearly with $|\delta \Psi|$ or $|\delta \Psi_\FN|$
\begin{eqnarray}
\delta O = {\cal O}(|\delta \Psi| + |\delta \Psi_\FN|),
\end{eqnarray}
but the use of the hybrid estimator ``$2\DMC-\VMC$'', $O_{\hybrid} = 2 \la O_L(\b{R}) \ra_{\Psi_\FN \Psi} - \la O_L(\b{R}) \ra_{\Psi^2}$, allows one to remove the dominant linear term $|\delta \Psi|$
\begin{eqnarray}
\delta O_{\hybrid} = {\cal O}(|\delta \Psi_\FN| + |\delta \Psi|^2 + |\delta \Psi_\FN|^2 + |\delta \Psi|\,|\delta \Psi_\FN|).
\nonumber\\
\label{deltaOhyb}
\end{eqnarray}

The same improved estimators defined in the previous section can also be used straightforwardly in FN-DMC calculations.
Note that, in this case, the average of the ZV term of Eq.~(\ref{DOLZV}) no longer vanishes on an infinite MC sample.
Nevertheless, this term gives the same order of reduction of the variance as in VMC. The systematic error and the variance
are still given by Eqs.~(\ref{deltaOZV}) and (\ref{sigma2OZV}), though the prefactors can be different in VMC and DMC.
We note that it is possible to define a new ZV correction of vanishing average in FN-DMC~\cite{AssCaf-JCP-00}
but this does not change the leading order of either the systematic error or the variance.

Adding the ZB term of Eq.~(\ref{DOLZB}), the systematic error of the ZVZB estimator vanishes quadratically
\begin{eqnarray}
\delta O^{\ZVZB} &=& {\cal O}(|\delta \Psi|^2 + |\delta \Psi|\,|\delta \Psi'| + |\delta \Psi|\,|\delta \Psi_\FN|
\nonumber\\
&&+ |\delta \Psi'|\,|\delta \Psi_\FN|),
\label{deltaOZVZBdmc}
\end{eqnarray}
with no linear term $|\delta \Psi_\FN|$ in contrast to the hybrid estimator of Eq.~(\ref{deltaOhyb}). The behavior of the variance of the ZVZB estimator in FN-DMC is still given by Eq.~(\ref{sigma2OZVZB}). One can also define a hybrid ZVZB estimator, $O^{\ZVZB}_{\hybrid} = 2 \la O_L^{\ZVZB}(\b{R}) \ra_{\Psi_\FN \Psi} - \la O_L^{\ZVZB}(\b{R}) \ra_{\Psi^2}$, whose systematic error also vanishes quadratically as
\begin{eqnarray}
\delta O^{\ZVZB}_{\hybrid} &=& {\cal O}(|\delta \Psi|^2 + |\delta \Psi|\,|\delta \Psi_\FN| + |\delta \Psi'|\,|\delta \Psi_\FN|),
\nonumber\\
\end{eqnarray}
with no term in $|\delta \Psi|\,|\delta \Psi'|$ in contrast to Eq.~(\ref{deltaOZVZBdmc}).

\subsection{Expressions of the estimators for local Hamiltonians}

We give now more explicit expressions of the estimators in terms of the convenient logarithmic derivative $Q(\b{R})=\Psi'(\b{R})/\Psi(\b{R})$.

In the case of local Hamiltonians $\bra{\b{R}} \hat{H} \ket{\b{R}'} = H(\b{R}) \delta(\b{R}-\b{R}')$ where $H(\b{R})= -(1/2) \sum_{k=1}^{N} \nabla_{\b{r}_k}^2 + V(\b{R})$ contains kinetic and potential energy contributions, the ZV term takes the form
\begin{eqnarray}
\Delta O_{L}^{\ZV}(\b{R}) &=& -\frac{1}{2} \sum_{k=1}^{N} \Big[ \nabla_{\b{r}_k}^2 Q(\b{R})
+ 2 \, \b{v}_k(\b{R}) \cdot \nabla_{\b{r}_k} Q(\b{R}) \Big],
\nonumber\\
\label{DOLZVHloc}
\end{eqnarray}
where $\b{v}_k(\b{R}) = \nabla_{\b{r}_k} \Psi(\b{R}) /\Psi(\b{R})$ is the drift velocity of the trial wave function. The ZB term is simply
\begin{eqnarray}
\Delta O_{L}^{\ZB}(\b{R}) = 2 \Delta E_L(\b{R}) Q(\b{R}),
\end{eqnarray}
where $\Delta E_L(\b{R}) = E_L(\b{R}) - E$. For each observable, a specific form for $Q(\b{R})$ has to be chosen. In principle, one could optimize $Q(\b{R})$ for each system by minimizing the variance of the ZVZB estimator. Besides $Q(\b{R})$, the ZV and ZB terms require only the evaluation of the drift velocity and the local energy of the trial wave function that are already available in QMC programs.

Clearly, a nonlocal operator $V(\b{R},\b{R}')$ in the Hamiltonian, e.g. a nonlocal pseudopotential, would yield an additional term in the simplified expression of the ZV correction of Eq.~(\ref{DOLZVHloc}). Although the inclusion of this additional term is required for the variance to rigorously vanish quadratically as in Eq.~(\ref{sigma2OZV}), in practice the estimator using the simple form of Eq.~(\ref{DOLZVHloc}) can already achieve a considerable variance reduction.

\section{Zero-variance zero-bias improved estimators for the spherically averaged position intracule density}
\label{sec:zvzb_intracule}

In the case of the spherically averaged position intracule density, the local observable $I_L(u,\b{R})$ is
\begin{equation}
I_L(u,\b{R}) = \frac{1}{2} \sum_{i \not= j}  \int \frac{d\Omega_{\b{u}}}{4\pi} \, \delta (\b{r}_{ij} -\b{u}),
\end{equation}
which has an infinite variance. We will give two possible choices for $Q(\b{R})$ which define good improved estimators.

\subsection{First improved estimator}

The minimal choice for $Q(\b{R})$ that cancels the infinite variance of $I_L(u,\b{R})$ is
\begin{equation}
Q_1(u,\b{R}) = - \frac{1}{8\pi} \sum_{i \not= j} \int \frac{d\Omega_{\b{u}}}{4\pi} \, \frac{1}{|\b{r}_{ij} -\b{u}|},
\label{Q1}
\end{equation}
which gives the following ZV contribution
\begin{eqnarray}
\Delta I_{L}^{\ZV1}(u,\b{R}) &=& \frac{1}{8\pi} \sum_{i \not= j} \int \frac{d\Omega_{\b{u}}}{4\pi} \, \Big[ \nabla_{\b{r}_{ij}}^2 \frac{1}{|\b{r}_{ij} -\b{u}|}
\nonumber\\
&& + 2 \,  \b{v}_i(\b{R})  \cdot  \nabla_{\b{r}_{ij}} \frac{1}{|\b{r}_{ij} -\b{u}|} \Big].
\end{eqnarray}
Using the identity $\nabla_{\b{r}_{ij}}^2 1/|\b{r}_{ij} -\b{u}| = - 4 \pi \delta (\b{r}_{ij} -\b{u})$, it is seen that the first term in $\Delta I_{L}^{\ZV1}(u,\b{R})$ exactly cancels the delta function in $I_{L}(u,\b{R})$, and the ZV estimator $I_{L}^{\ZV1}(u,\b{R}) = I_{L}(u,\b{R}) + \Delta I_{L}^{\ZV1}(u,\b{R})$ is simply
\begin{eqnarray}
I_{L}^{\ZV1}(u,\b{R}) &=&  \frac{1}{4\pi} \sum_{i \not= j} \b{v}_i(\b{R}) \cdot  \int \frac{d\Omega_{\b{u}}}{4\pi} \,  \nabla_{\b{r}_{ij}} \frac{1}{|\b{r}_{ij} -\b{u}|}
\nonumber\\
&=& - \frac{1}{4\pi} \sum_{i \not= j} \b{v}_i(\b{R}) \cdot  \frac{\b{r}_{ij}}{r_{ij}^3} \, 1_{[u,+\infty[}(r_{ij}). \, \,
\label{IZV1}
\end{eqnarray}
The form of $Q_1(u,\b{R})$ of Eq.~(\ref{Q1}) could also have been guessed from the behavior at small $\b{r}_{ij}$ of the first-order solution $\Psi'(\b{R})$ of the perturbation theory with respect to $\delta(\b{r}_{ij} -\b{u})$ (see Refs.~\onlinecite{Sch-AP-59,Tri-JPB-80}). In contrast to the histogram estimator of Eq.~(\ref{esthisto}) where only electron pairs with relative distance $r_{ij}$ in the small interval $[u-\Delta u/2, u+\Delta u/2[$ contribute to $I(u)$, for the ZV estimator of Eq.~(\ref{IZV1}) all the electron pairs with relative distance $r_{ij} \geq u$ contribute to $I(u)$. For $u \not = 0$, the variance of this ZV estimator is finite and much smaller than the variance of the histogram estimator as shown in Sec.~\ref{sec:results}. For $u = 0$, this ZV estimator reduces to the estimator proposed in Ref.~\onlinecite{LanRotVrb-JCP-97} to compute one-electron densities at the nucleus and applied in Ref.~\onlinecite{SarGalBue-JCP-98} for calculations of intracule densities at the coalescence point.

The ZB correction corresponding to $Q_1(u,\b{R})$ of Eq.~(\ref{Q1}) is
\begin{eqnarray}
\Delta I_{L}^{\ZB1}(u,\b{R}) = - \frac{\Delta E_L(\b{R})}{4\pi} \sum_{i \not= j}  \int \frac{d\Omega_{\b{u}}}{4\pi} \, \frac{1}{|\b{r}_{ij} -\b{u}|}
\nonumber\\
= - \frac{\Delta E_L(\b{R})}{4\pi} \sum_{i \not= j} \Big[ \frac{1_{[0,u[}(r_{ij})}{u} + \frac{1_{[u,+\infty[}(r_{ij})}{r_{ij}}  \Big].
\label{DIZB1}
\end{eqnarray}
The resulting ZVZB estimator $I_{L}^{\ZV1\ZB1}(u,\b{R}) = I_{L}^{\ZV1}(u,\b{R}) +  \Delta I_{L}^{\ZB1}(u,\b{R})$ allows one to reduce the systematic error due to the trial wave function as shown in Sec.~\ref{sec:results}. Note that, in contrast to the histogram estimator, these ZV and ZVZB improved estimators do not suffer from any discretization error: they are estimators of $I(u)$ for an \textit{infinitely precise} value of $u$. Both the ZV estimator of Eq.~(\ref{IZV1}) and its ZB correction of Eq.~(\ref{DIZB1}) are very simple to program and very fast to compute on a grid over $u$.

Like the histogram estimator, the ZV estimator of Eq.~(\ref{IZV1}) gives simply zero for distances $u$ beyond
the largest electron-electron distance $r_{ij}$ encountered in the Monte Carlo run.
Also, because the form of $Q_1(u,\b{R})$ of Eq.~(\ref{Q1}) has been chosen only to cure the deficiency of the histogram
estimator at small $u$, the ZB correction of Eq.~(\ref{DIZB1}) actually makes the histogram estimator worse at large $u$.
The ZB correction decays only as $1/u$ as $u \to \infty$ instead of the correct exponential decay and
consequently gives large variances at large $u$. Thus, it is inaccurate to extract the long-range behavior of the
intracule density using the ZVZB estimator $I_{L}^{\ZV1\ZB1}(u,\b{R})$.
The second choice for $Q(\b{R})$ presented next remedies this limitation.

\subsection{Second improved estimator}
\label{second_estimator}

A more general choice for $Q(\b{R})$ that cancels the infinite variance of $I_L(u,\b{R})$ is
\begin{equation}
Q_2(u,\b{R}) = - \frac{1}{8\pi} \sum_{i \not= j} \int \frac{d\Omega_{\b{u}}}{4\pi} \, \frac{g(\b{r}_{ij},\b{u})}{|\b{r}_{ij} -\b{u}|},
\end{equation}
where $g(\b{r}_{ij},\b{u})$ is some function satisfying $g(\b{r}_{ij},\b{u})=1$ for $\b{r}_{ij}=\b{u}$. For example, using
\begin{equation}
g(\b{r}_{ij},\b{u}) = e^{-\z |\b{r}_{ij} - \b{u}|},
\end{equation}
with $\z> 0$, has the advantage of ensuring a correct exponential decay of the estimator at infinity for a finite system.

The corresponding ZV estimator is
\begin{eqnarray}
I_{L}^{\ZV2}(u,\b{R}) &=& \frac{1}{8\pi} \sum_{i \not= j} \int \frac{d\Omega_{\b{u}}}{4\pi} \, \Big[ 2 \,
\b{v}_i(\b{R}) \cdot  \nabla_{\b{r}_{ij}} \frac{e^{-\z |\b{r}_{ij} - \b{u}|}}{|\b{r}_{ij} -\b{u}|}
\nonumber\\
&&+ 2 \, \nabla_{\b{r}_{ij}} \frac{1}{|\b{r}_{ij} -\b{u}|} \cdot \nabla_{\b{r}_{ij}} e^{-\z |\b{r}_{ij} - \b{u}|}
\nonumber\\
&&+  \frac{1}{|\b{r}_{ij} -\b{u}|} \nabla_{\b{r}_{ij}}^2 e^{-\z |\b{r}_{ij} - \b{u}|} \Big]
\nonumber\\
&=& -\frac{1}{4\pi} \sum_{i \not= j} \, \Big[ \b{v}_i(\b{R}) \cdot \frac{\b{r}_{ij}}{r_{ij}^3} \Big( r_{ij} A(r_{ij},u)
\nonumber\\
&& + \frac{r_{ij}}{u} B(r_{ij},u) \Big)
- \frac{\z^2}{2} A(r_{ij},u) \Big],
\label{IZV2}
\end{eqnarray}
and its ZB correction is
\begin{eqnarray}
\Delta I_{L}^{\ZB2}(u,\b{R}) &=& - \frac{\Delta E_L(\b{R})}{4\pi} \sum_{i \not= j}  \int \frac{d\Omega_{\b{u}}}{4\pi} \, \frac{e^{-\z |\b{r}_{ij} - \b{u}|}}{|\b{r}_{ij} -\b{u}|}
\nonumber\\
&=& - \frac{\Delta E_L(\b{R})}{4\pi} \sum_{i \not= j} A (r_{ij},u),
\label{DIZB2}
\end{eqnarray}
where
\begin{eqnarray}
A(r_{ij},u) =  \frac{e^{-\z|r_{ij}-u|} - e^{-\z(r_{ij}+u)}}{2 \z r_{ij} u},
\end{eqnarray}
\begin{eqnarray}
B(r_{ij},u) = \frac{1}{2} \left[ \sgn(r_{ij}-u) e^{-\z|r_{ij}-u|} - e^{-\z(r_{ij}+u)} \right],
\end{eqnarray}
and
where $\sgn$ is the sign function.
These refined ZV estimator $I_{L}^{\ZV2}(u,\b{R})$ and ZVZB estimator $I_{L}^{\ZV2\ZB2}(u,\b{R}) = I_{L}^{\ZV2}(u,\b{R}) + \Delta I_{L}^{\ZB2}(u,\b{R})$ now both decay correctly exponentially as $u\to\infty$. The constant $\z$ is chosen according to Eq.~(\ref{Idecay}), i.e. $\z=2 \sqrt{2 I}$ where $I$ is an estimate of the vertical ionization energy.
On the other hand, the computational cost is larger per Monte Carlo step than that of the simpler estimators of the previous subsection.

\section{Computational details}
\label{sec:compdetails}

We have chosen to illustrate the efficiency and accuracy of the new improved QMC estimators by calculating the intracule density of the He and C atoms and the C$_2$ and N$_2$ molecules in their ground states. The molecules are considered at their experimental geometries ($d_{\C-\C}=2.3481$ Bohr~\cite{CadWah-ADNDT-74} and $d_{\N-\N}=2.075$ Bohr~\cite{HubHer-BOOK-79}).

We start by generating a standard {\it ab initio} wave function using the quantum chemistry program GAMESS~\cite{SchBalBoaElbGorJenKosMatNguSuWinDupMon-JCC-93}, typically a HF wave function or a MCSCF wave function with a complete active space (CAS) generated by distributing $n$ valence electrons in $m$ valence orbitals [CAS($n$,$m$)]. For each system considered, we choose the one-electron Slater basis so as to ensure that the HF intracule density is reasonably converged with respect to the basis. For the He atom, we use the basis of Clementi and Roetti~\cite{CleRoe-ADNDT-74}, with exponents reoptimized at the HF level by Koga {\it et al.}~\cite{KogTatTha-PRA-93}. For the C atom, we use the CVB1 basis of Ema {\it et al.}~\cite{EmaGarRamLopFerMeiPal-JCC-03}. For the C$_2$ and N$_2$ molecules, we use the CVB2 basis of the same authors. In GAMESS, each Slater function is actually approximated by a fit to 14 Gaussian functions~\cite{HehStePop-JCP-69,Ste-JCP-70,KolReiAss-JJJ-XX}

This standard {\it ab initio} wave function is then multiplied by a Jastrow factor, imposing the electron-electron cusp condition, but with essentially all other free parameters chosen to be zero to form our starting trial Jastrow-Slater wave function, and QMC calculations are performed with the program CHAMP~\cite{Cha-PROG-XX} or QMCMOL~\cite{Qmc-PROG-XX} using the true Slater basis set rather than its Gaussian expansion. The Jastrow parameters, the orbital coefficients and the configuration state functions (CSFs) coefficients of this trial wave function are optimized in VMC using the very efficient, recently-developed linear energy minimization method~\cite{TouUmr-JCP-07,UmrTouFilSorHen-PRL-07} and an accelerated Metropolis algorithm~\cite{Umr-PRL-93,Umr-INC-99}. Once the trial wave function has been optimized, we compute the intracule density in VMC, and in DMC using the fixed-node and the short-time approximations (see, e.g., Refs.~\onlinecite{And-JCP-75,And-JCP-76,ReyCepAldLes-JCP-82,MosSchLeeKal-JCP-82}). We use an imaginary time-step of $\tau=0.01$ hartree$^{-1}$ in an efficient DMC algorithm featuring very small time-step errors~\cite{UmrNigRun-JCP-93}.

\section{Results and discussion}
\label{sec:results}

\subsection{He atom}

The improvement due to the ZV and ZVZB estimators is illustrated for the simple case of He atom.

\begin{figure}
\includegraphics[scale=0.35,angle=-90]{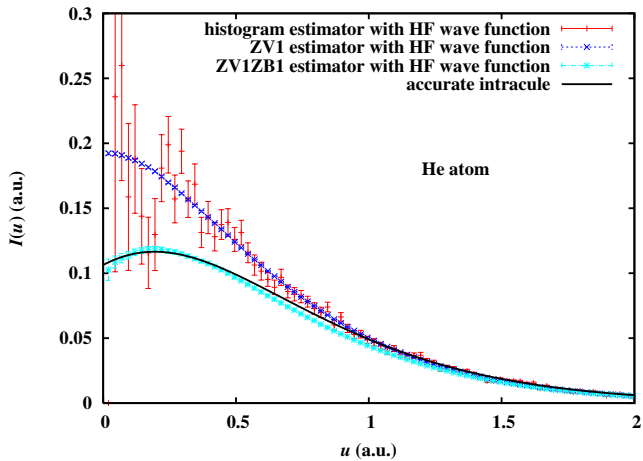}
\caption{Spherically averaged position intracule density $I(u)$ as a function of the electron-electron distance $u$ for the He atom. The intracule densities calculated in VMC with the histogram, ZV1 and ZV1ZB1 estimators using only 100~000 MC configurations sampled from a HF trial wave function (without a Jastrow factor) are compared. For the histogram estimator, a grid step of $\Delta u = 0.005$ has been used. The intracule density obtained with an accurate, explicitly-correlated Hylleraas-type wave function is also shown.
}
\label{fig:he_ktt_hf_vmc_int}
\end{figure}

Figure~\ref{fig:he_ktt_hf_vmc_int} shows the spherically averaged intracule density $I(u)$ as a function of the electron-electron distance $u$, calculated in VMC using the standard histogram estimator of Eq.~(\ref{esthisto}), the ZV1 improved estimator of Eq.~(\ref{IZV1}) and the ZV1ZB1 improved estimator of Eq.~(\ref{DIZB1}), employing only 100~000 MC configurations sampled from a HF trial wave function (without a Jastrow factor).
In this subsection, we intentionally use a poor trial wave function to demonstrate the large reduction in the bias resulting from the ZVZB estimator.
The calculations take a few seconds on a present-day single-processor personal computer. The intracule density obtained with an accurate, explicitly-correlated Hylleraas-type wave function with 491 terms~\cite{FreHuxMor-PRA-84,UmrGon-PRA-94,CioSteTanUmr-JCP-95} is also shown as a reference. The statistical uncertainty obtained with the histogram estimator is very large, especially at small $u$, making it impossible to extract the on-top intracule density $I(0)$. The ZV1 estimator spectacularly reduces the statistical uncertainty which becomes invisible at the scale of the plot, its variance being smaller by nearly 4 orders of magnitude for small and large $u$ and by about $2$ orders of magnitude for intermediate $u$. The ZV1ZB1 estimator in turn spectacularly reduces the systematic error due to the use of a HF trial wave function, the obtained intracule density agreeing well with the accurate reference. In particular, the maximum at small $u$ caused by the cusp at $u=0$ is accurately reproduced. On the other hand, the use of the ZV1ZB1 estimator increases slightly the variance at small $u$ and more importantly at large $u$ where the ZB1 correction of Eq.~(\ref{DIZB1}) decays too slowly as $1/u$.

\begin{figure}
\includegraphics[scale=0.35,angle=-90]{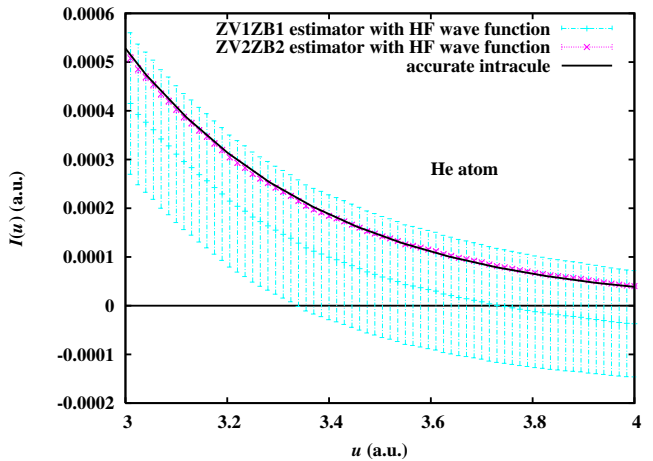}
\caption{Long-range tail of the spherically averaged position intracule density $I(u)$ of the He atom. The intracule densities calculated in VMC with the ZV1ZB1 and ZV2ZB2 (with $\zeta = 2\sqrt{2I} \approx 2.7$) estimators using only 100~000 MC configurations sampled from a HF trial wave function (without a Jastrow factor) are compared. The intracule density obtained with an accurate, explicitly-correlated Hylleraas-like wave function is also shown.
}
\label{fig:he_ktt_hf_vmc_int_zvzb}
\end{figure}

The behavior of the estimators in the long-range tail of $I(u)$ is explored in detail in Fig.~\ref{fig:he_ktt_hf_vmc_int_zvzb} which compares the ZV1ZB1 and ZV2ZB2 estimators of Eqs.~(\ref{IZV2}) and~(\ref{DIZB2}) for $u>3$. As mentioned in Sec.~\ref{second_estimator}, the relative statistical uncertainty obtained with the ZV1ZB1 is very large (about $200\%$ at $u=3.5$), whereas the ZV2ZB2 estimator, which has the correct exponential decay at large $u$, spectacularly reduces the statistical uncertainty, its variance being smaller by about 3 orders of magnitude at $u=3.5$. The improvement is even more dramatic for larger $u$.

We now discuss the scaling of the intracule density and its statistical uncertainty with respect to the nuclear charge $Z$. For this purpose, we have calculated the intracule densities of the elements of the He isoelectronic series from $Z=2$ (He) to $Z=20$ (Ar$^{18+}$) using Jastrow-Slater wave functions. We found that, in the relevant range of $u$ ($u \lesssim 5$), the intracule density roughly obeys the scaling law $I(u/Z) \sim Z^{n}$ with $n\approx 3$, in agreement with the value predicted by a simple hydrogenic model~\cite{KogMat-JCP-97}. The statistical uncertainty of $I(u/Z)$ computed with the ZV1 estimator roughly obeys the same law, so that the relative statistical accuracy on $I(u/Z)$ does not deteriorate with $Z$ in the series (at least up to $Z=20$).

Finally, we note in passing that Fig.~\ref{fig:he_ktt_hf_vmc_int_zvzb} also reveals another advantage of our improved estimators: because the estimator of $I(u)$ is strongly correlated with the estimator of $I(u +\Delta u)$, as obvious for instance in Eq.~(\ref{IZV1}), the obtained intracule density curves are always very smooth, even at a scale much smaller than the statistical uncertainty. This makes the calculated intracule densities directly suitable for subsequent manipulations.

\subsection{C atom}

We continue the discussion of the improved estimators for the more interesting case of the C atom.

\begin{figure}
\includegraphics[scale=0.35,angle=-90]{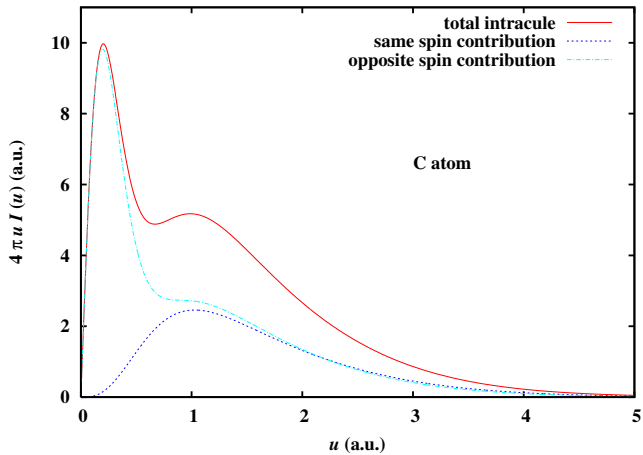}
\caption{Integrand of the electron-electron Coulomb interaction energy $4\pi u \, I(u)$ over the electron-electron distance $u$ for the C atom, calculated in VMC using the ZVZB2 (with $\zeta = 2\sqrt{2I} \approx 1.82$~\cite{Nis-BOOK-05}) improved estimator and a fully optimized Jastrow-Slater CAS(4,4) trial wave function. The same spin contribution $4\pi u \, I^{\SS}(u)$ and the opposite spin contribution $4\pi u \, I^{\OS}(u)$ are also shown.
}
\label{fig:c_cvb1_cas44j_vmc_emin_csfoj_4piui_zvzb5}
\end{figure}

The repercussion of the shell structure of the system on the intracule is apparent in Fig.~\ref{fig:c_cvb1_cas44j_vmc_emin_csfoj_4piui_zvzb5} which shows the integrand of the electron-electron Coulomb interaction energy over the electron-electron distance $u$, i.e.  $4\pi u \, I(u)$ [see Eq.~(\ref{Wee})]. The ZVZB2 estimator is used with an accurate Jastrow-Slater CAS(4,4) wave function in which the Jastrow, CSF and orbital parameters have been simultaneously optimized. Note however that, at the scale of the plot, the HF intracule $4\pi u \, I_\HF(u)$ would look the same. The curve displays two maxima: the maximum at short distance ($u \approx 0.2$) corresponds essentially to the K-K electron pair and the maximum at longer distance ($u \approx 1.0$) corresponds essentially to the K-L and L-L electron pairs. Fig.~\ref{fig:c_cvb1_cas44j_vmc_emin_csfoj_4piui_zvzb5} also shows the same spin (SS) contribution
\begin{eqnarray}
I^{\SS}(u) = I^{\uparrow\uparrow}(u) + I^{\downarrow\downarrow}(u),
\end{eqnarray}
and opposite spin (OS) contribution
\begin{eqnarray}
I^{\OS}(u) = I^{\uparrow\downarrow}(u) + I^{\downarrow\uparrow}(u),
\end{eqnarray}
where $I^{\sigma\sigma'}(u)$ is the spherical average of the spin-resolved intracule densities
\begin{eqnarray}
I^{\sigma\sigma'}(\b{u}) = \frac{1}{2} \sum_{i \not= j} \int d\b{R} \, \Psi(\b{R})^2 \, \delta(\b{r}_{ij} - \b{u}) \delta_{s_i,\sigma} \delta_{s_j,\sigma'},
\end{eqnarray}
where the spin coordinates of the $N=N_{\uparrow}+N_{\downarrow}$ electrons in the spin-assigned wave function $\Psi(\b{R})$ are fixed at $s_i=\uparrow$ for $i=1,...,N_{\uparrow}$ and $s_i=\downarrow$ for $i=N_{\uparrow}+1,..., N_{\uparrow}+N_{\downarrow}$. Not surprisingly, the intracule density at short distance is dominated by the OS contribution, while at long distance the SS and OS contributions become nearly identical.

\begin{figure}
\includegraphics[scale=0.35,angle=-90]{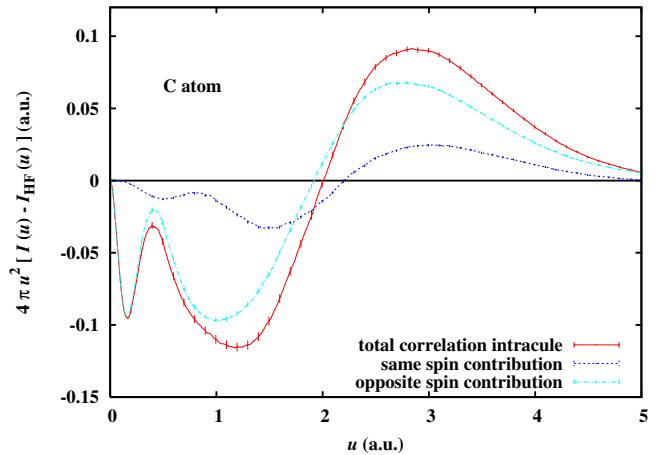}
\caption{Correlation part of the radial position intracule density $4\pi u^2 \left[ I(u) - I_{\HF}(u)\right]$ as a function of the electron-electron distance $u$ for the C atom, where $I(u)$ has been calculated in VMC using the ZVZB2 (with $\zeta = 2\sqrt{2I} \approx 1.82$~\cite{Nis-BOOK-05}) improved estimator and a fully optimized Jastrow-Slater CAS(4,4) trial wave function. The same spin contribution $4\pi u^2 \, \left[I^{\SS}(u) - I^{\SS}_{\HF}(u)\right]$ and the opposite spin contribution $4\pi u^2 \, \left[I^{\OS}(u) - I^{\OS}_{\HF}(u)\right]$ are also shown.
}
\label{fig:c_cvb1_cas44j_vmc_emin_csfoj_4piu2ic_zvzb5}
\end{figure}

Figure~\ref{fig:c_cvb1_cas44j_vmc_emin_csfoj_4piu2ic_zvzb5} shows the correlation part of the radial intracule density $4\pi u^2 \left[ I(u) - I_{\HF}(u)\right]$, also called correlation hole, using the ZVZB2 improved estimator and the same Jastrow-Slater CAS(4,4) trial wave function. Correlation decreases the radial probability density at the two previously-mentioned maxima and increases it at longer distances. The same spin contribution $4\pi u^2 \, \left[I^{\SS}(u) - I^{\SS}_{\HF}(u)\right]$ and the opposite spin contribution $4\pi u^2 \, \left[I^{\OS}(u) - I^{\OS}_{\HF}(u)\right]$ are also shown in Fig.~\ref{fig:c_cvb1_cas44j_vmc_emin_csfoj_4piu2ic_zvzb5}. The OS component constitutes the main contribution to the correlation hole.

\begin{figure}
\includegraphics[scale=0.35,angle=-90]{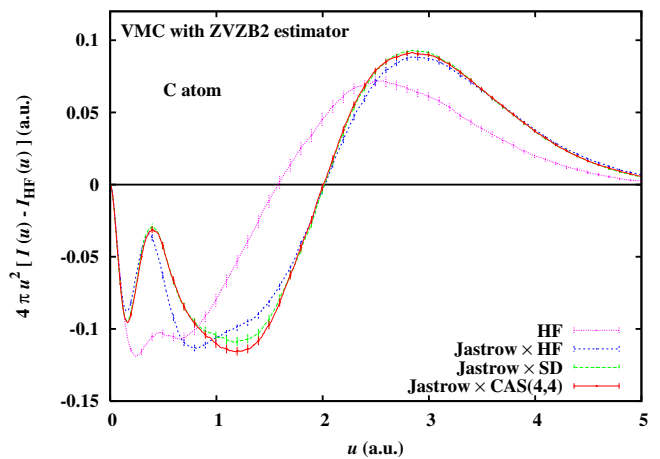}
\caption{Correlation part of the radial position intracule density $4\pi u^2 \left[ I(u) - I_{\HF}(u)\right]$ as a function of the electron-electron distance $u$ for the C atom, where $I(u)$ has been calculated in VMC using the ZVZB2 (with $\zeta = 2\sqrt{2I} \approx 1.82$~\cite{Nis-BOOK-05}) improved estimator for a series of trial wave functions of increasing accuracy: HF, Jastrow $\times$ HF, fully-optimized Jastrow $\times$ SD and Jastrow $\times$ CAS(4,4) wave functions.
}
\label{fig:c_cvb1_vmc_emin_4piu2ic_zvzb5}

\end{figure}

The effect of the chosen trial wave function on the accuracy on the correlation hole is examined in Fig.~\ref{fig:c_cvb1_vmc_emin_4piu2ic_zvzb5}. Four trial wave functions of increasing accuracy have been tested:  HF, Jastrow $\times$ HF (with only Jastrow parameters optimized), Jastrow $\times$ single-determinant (SD) (with Jastrow and orbital parameters optimized) and Jastrow $\times$ CAS(4,4) (with Jastrow, CSF and orbital parameters optimized). Remarkably, the ZVZB2 estimator gives a correlation hole with a correct overall structure even with the uncorrelated HF wave function. For more quantitative predictions, the use of a Jastrow factor is however necessary. Optimization of the orbitals in a Jastrow-Slater single-determinant wave function brings a further significant improvement in the interesting short-range region ($u \lesssim 2$).
The Jastrow-Slater multi-determinant CAS(4,4) and the Jastrow-Slater single-determinant intracules agree closely.

\subsection{C$_2$ molecule}

We now discuss the more difficult case of the C$_2$ molecule. The ground-state wave function of this system has a strong multi-configurational character due to energetic near-degeneracies among the valence orbitals (``nondynamical correlation'' in chemists' jargon, or ``strong correlation'' in physicists' jargon), making it a challenging system despite its small size.

\begin{figure}
\includegraphics[scale=0.35,angle=-90]{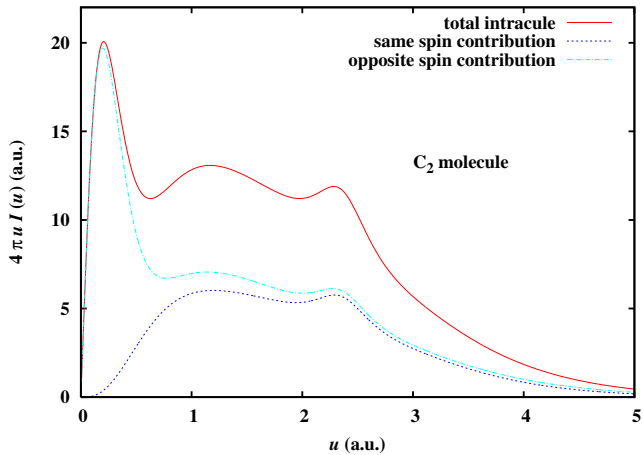}
\caption{Integrand of the electron-electron Coulomb interaction energy $4\pi u \, I(u)$ versus the electron-electron distance $u$ for the C$_2$ molecule, calculated in VMC using the ZVZB2 (with $\zeta = 2\sqrt{2I} \approx 1.83$~\cite{Nis-BOOK-05}) improved estimator and a fully optimized Jastrow-Slater CAS(8,8) trial wave function. The same spin contribution $4\pi u \, I^{\SS}(u)$ and the opposite spin contribution $4\pi u \, I^{\OS}(u)$ are also shown.
}
\label{fig:c2_cvb2_cas88j_vmc_emin_csfoj_4piui_zvzb5}
\end{figure}

Figure~\ref{fig:c2_cvb2_cas88j_vmc_emin_csfoj_4piui_zvzb5} plots $4\pi u \, I(u)$ versus $u$ using the ZVZB2 estimator with an accurate Jastrow-Slater CAS(8,8) wave function (Jastrow, CSF and orbital parameters optimized). The curve displays three maxima: the maximum at short distance ($u \approx 0.2$) corresponds essentially to the intra-atomic K-K electron pairs; the maximum at long distance ($u \approx 2.3$) is located at about the interatomic distance and corresponds mainly to the interatomic K-K electron pairs; the maximum at intermediate distance ($u \approx 1.2$) must correspond mainly to intra-atomic and interatomic K-L and L-L electron pairs. As for the C atom, the intracule density at short distance is dominated by the OS contribution, while at long distance the SS and OS contributions become nearly identical.

\begin{figure}
\includegraphics[scale=0.35,angle=-90]{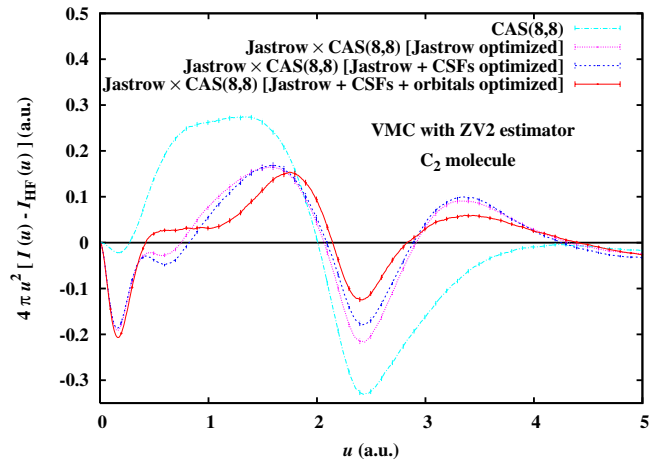}
\caption{Correlation part of the radial position intracule density $4\pi u^2 \left[ I(u) - I_{\HF}(u)\right]$ as a function of the electron-electron distance $u$ for the C$_2$ molecule, where $I(u)$ has been calculated in VMC using the ZV2 improved estimator (with $\zeta = 2\sqrt{2I} \approx 1.83$~\cite{Nis-BOOK-05}) with a MCSCF CAS(8,8) wave function and a series of Jastrow $\times$ CAS(8,8) wave functions with different levels of optimization.
}
\label{fig:c2_cvb2_cas88_vmc_4piu2ic_zv5}
\end{figure}

Figure~\ref{fig:c2_cvb2_cas88_vmc_4piu2ic_zv5} shows the correlation hole of the C$_2$ molecule using the ZV2 improved estimator calculated in VMC with a MCSCF CAS(8,8) wave function and a series of Jastrow $\times$ CAS(8,8) wave functions for different levels of optimization. The MCSCF CAS(8,8) wave function does not correlate the core electrons and thus gives essentially no correlation hole at distances corresponding to intra-atomic core electron pairs ($u \approx 0.2$). At longer valence distances, the MCSCF CAS(8,8) wave function gives only a gross overall shape of the correlation hole. Introduction of a Jastrow factor yields a correct correlation hole at core distances, which as expected is about twice as deep as the core correlation hole of the C atom, and reduces the correlation hole at valence distances. The action of the Jastrow factor is thus not limited to very short electron-electron distances, but in fact importantly modifies the correlation hole up to distances $u \approx$ 4. The Jastrow factor has also an indirect action via reoptimization of the CSF and orbital coefficients in its presence. The reoptimization of the CSF coefficients changes slightly the correlation hole at valence distances. The reoptimization of the CSF and orbital coefficients has a more important impact of the correlation hole at valence distances and also at core distances.

\begin{figure*}
\includegraphics[scale=0.35,angle=-90]{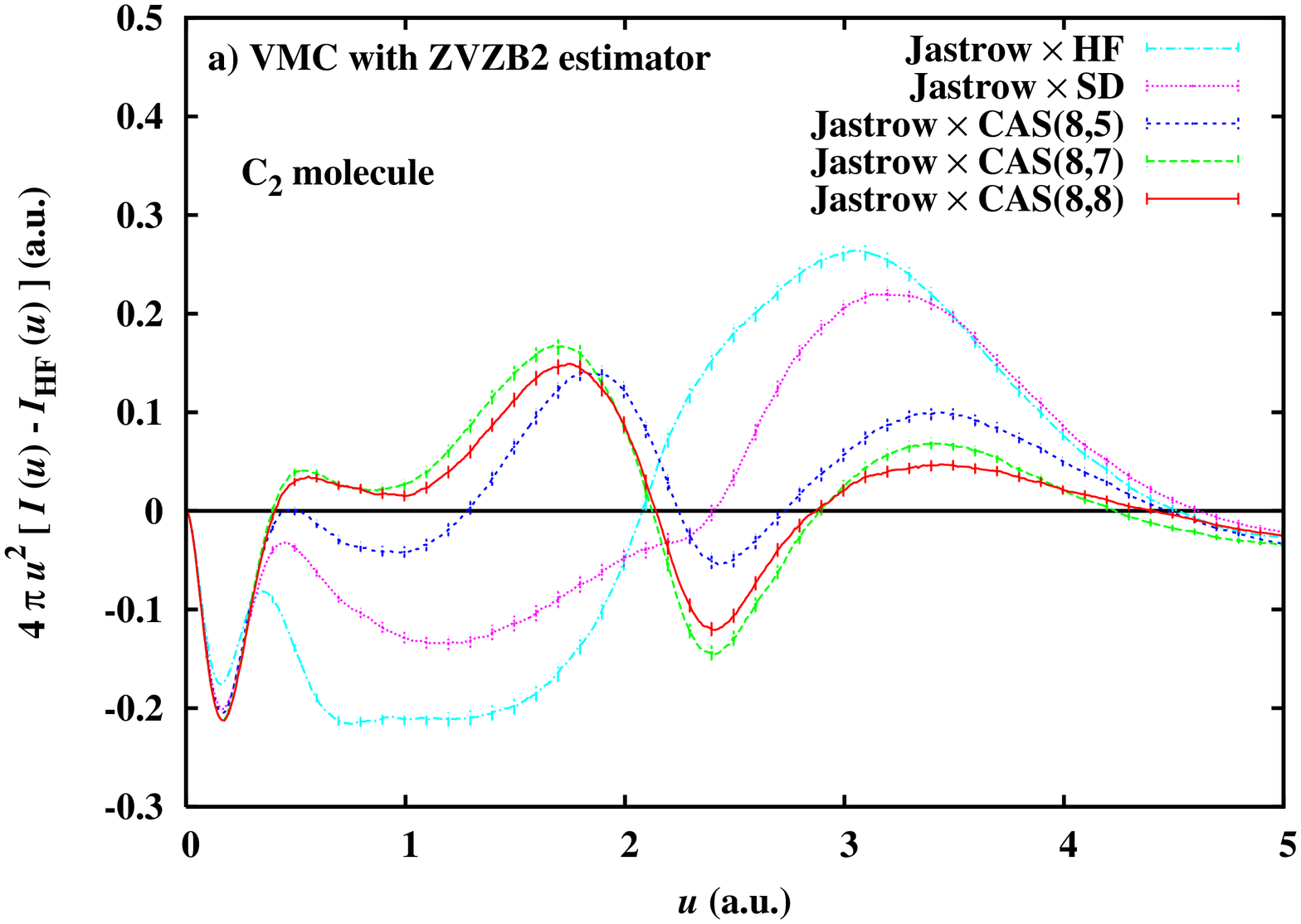}
\includegraphics[scale=0.35,angle=-90]{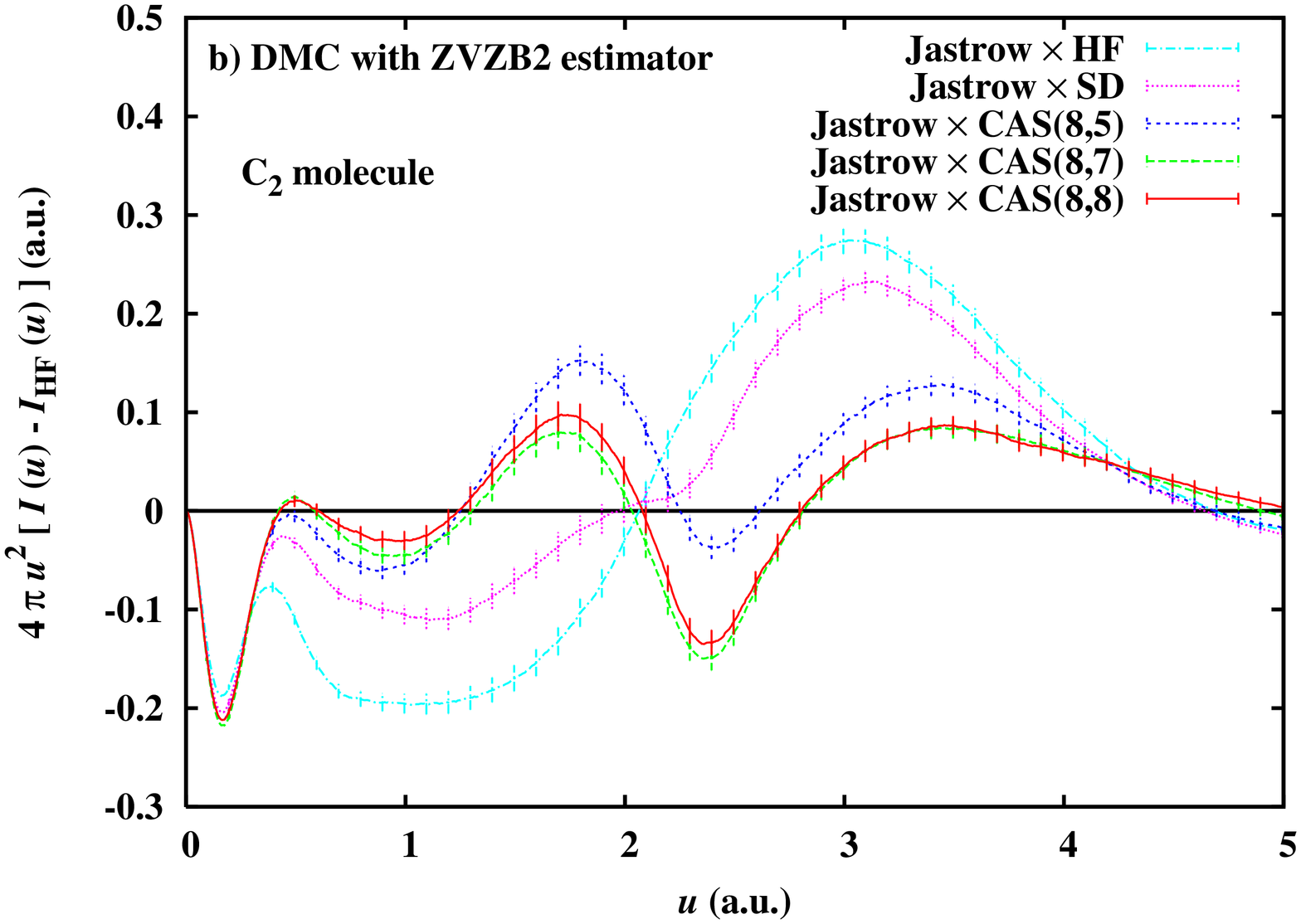}
\caption{Correlation part of the radial position intracule density $4\pi u^2 \left[ I(u) - I_{\HF}(u)\right]$ as a function of the electron-electron distance $u$ for the C$_2$ molecule, where $I(u)$ has been calculated in VMC (a) and FN-DMC (b) using the ZVZB2 (with $\zeta = 2\sqrt{2I} \approx 1.83$~\cite{Nis-BOOK-05}) improved estimator for a series of trial wave functions of increasing accuracy: Jastrow $\times$ HF, fully-optimized Jastrow $\times$ SD, Jastrow $\times$ CAS(8,5), Jastrow $\times$ CAS(8,7) and Jastrow $\times$ CAS(8,8) wave functions.
}
\label{fig:c2_cvb2_4piu2ic_zvzb5}
\end{figure*}

We next scrutinize in detail the convergence of the intracule density with respect to the determinantal part of the trial wave function. Figure~\ref{fig:c2_cvb2_4piu2ic_zvzb5}a shows the correlation hole using the ZVZB2 improved estimator calculated in VMC with five trial Jastrow-Slater wave functions of increasing accuracy: Jastrow $\times$ HF (with only the Jastrow parameters optimized), Jastrow $\times$ SD (with the Jastrow and orbital parameters optimized), and multi-determinant Jastrow $\times$ CAS(8,5), Jastrow $\times$ CAS(8,7) and Jastrow $\times$ CAS(8,8) (with the Jastrow, CSF and orbital parameters optimized). At core distances ($u \approx 0.2$), the correlation hole is essentially converged with only a single-determinant Jastrow-Slater wave function, provided that the orbitals are reoptimized together with the Jastrow factor. In contrast, at longer valence distances, the correlation hole depends very strongly on the determinantal part of the wave function. In particular, the multi-determinant wave functions yield a depletion of the intracule density at about the bond length, a feature not present at the single-determinant level. We have verified that this minimum disappears when using a pseudopotential removing the $1s$ electrons, and we thus interpret it as a decrease of probability of finding two interatomic core electrons separated by the bond distance. Overall, it appears necessary to use at least a multi-determinant CAS(8,7) wave function which includes configurations constructed from the antibonding $\pi$ orbitals to reach reasonable convergence. We have checked that further excitations beyond the CAS(8,8) wave function barely change the correlation hole. The corresponding correlation holes calculated in FN-DMC with the same ZVZB2 estimator and trial wave functions are reported in Fig.~\ref{fig:c2_cvb2_4piu2ic_zvzb5}b. The DMC intracules do not differ much from the corresponding VMC intracules, the accuracy of the intracule densities in the valence region being still essentially controlled by the trial wave function.

\subsection{N$_2$ molecule}

We finish our illustration of the calculation of intracule densities with the N$_2$ molecule, which has recently been the object of experimental and theoretical investigations~\cite{WatKamYamUdaMul-MP-04}.

\begin{figure}
\includegraphics[scale=0.35,angle=-90]{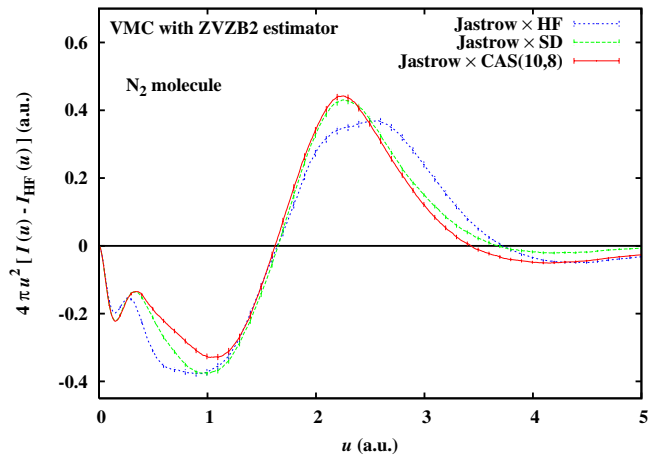}
\caption{Correlation part of the radial position intracule density $4\pi u^2 \left[ I(u) - I_{\HF}(u)\right]$ as a function of the electron-electron distance $u$ for the N$_2$ molecule, where $I(u)$ has been calculated in VMC using the ZVZB2 (with $\zeta = 2\sqrt{2I} \approx 2.14$~\cite{Nis-BOOK-05}) improved estimator for a series of trial wave functions of increasing accuracy: Jastrow $\times$ HF, fully-optimized Jastrow $\times$ SD and Jastrow $\times$ CAS(10,8) wave functions.
}
\label{fig:n2_cvb2_vmc_4piu2ic_zvzb5}
\end{figure}

The correlation holes of this molecule calculated in VMC with the ZVZB2 improved estimator for a Jastrow $\times$ HF, and fully optimized Jastrow $\times$ SD and Jastrow $\times$ CAS(10,8) wave functions are plotted in Fig.~\ref{fig:n2_cvb2_vmc_4piu2ic_zvzb5}. In sharp contrast to the C$_2$ molecule, no depletion of intracule density is observed at the bond length.
Also, the correlation hole is here essentially converged with only a single-determinant Jastrow-Slater wave function provided that the orbitals are optimized with the Jastrow factor. Unlike the C$_2$ molecule, going to a multi-determinant wave function does not have a large effect on the correlation hole at valence distances. The correlation hole calculated here agrees reasonably well with recent multi-reference configuration interaction and coupled-cluster calculations, and with experiment~\cite{WatKamYamUdaMul-MP-04}.

\section{Conclusions}
\label{sec:conclusion}

We have presented improved QMC estimators for the spherically averaged position intracule density $I(u)$, constructed using the general zero-variance zero-bias principle for observables that do not commute with the Hamiltonian. By replacing the average of the local delta-function operator by the average of a smooth non-local operator, these estimators decrease the variance of the standard histogram estimator by several orders of magnitude, and thus make the calculation of this quantity in QMC vastly more efficient. Interestingly, they permit calculations of $I(u)$ for very short and very large interelectronic distances $u$ that are never realized in the Monte Carlo run. These new estimators can also decrease the systematic error of the intracule density due to the approximate trial wave function. Other advantages of these estimators are the absence of any discretization error with respect to $u$ and the possibility to obtain very smooth curves for $I(u)$. These improved estimators, together with the achievement of systematically reducing the systematic error in both VMC and DMC calculations by optimization of trial wave functions with an increasing number of parameters, have allowed us to obtain accurate correlated intracule densities for atoms and molecules.

The estimators presented here can be used with trivial adaptations for QMC calculations of the companion entity of $I(u)$, namely the spherically averaged extracule density $E(r)$, representing the probability density of finding two electrons with center of mass at a radial distance $r$ with respect to the chosen origin. Similar improved estimators can be constructed for the three-dimensional intracule density $I(\b{u})$, extracule density $E(\b{r})$ and the full pair density $n_{2}(\b{r}_1,\b{r}_2)$. More generally, the variance reduction technique presented here can be applied to calculations of any pair-correlation function in classical and quantum Monte Carlo calculations.

\begin{acknowledgments}
We would like to thank Andreas Savin and Paola Gori-Giorgi for stimulating discussions. We also thank Alexander Kollias for providing us the Gaussian fits of Slater basis functions of Ref.~\onlinecite{KolReiAss-JJJ-XX}. J. T. acknowledges financial support from a Marie Curie Outgoing International Fellowship (039750-QMC-DFT). This work was also supported by the Centre National de la Recherche Scientifique and by the National Science Foundation (DMR-0205328, EAR-0530301).
Most of the calculations were been performed at the Cornell Theory Center and on the Intel cluster at the Cornell Nanoscale Facility (a member of the National Nanotechnology Infrastructure Network supported by the National Science Foundation).
\end{acknowledgments}

\bibliographystyle{apsrev}
\bibliography{biblio}

\end{document}